\documentclass[a4paper]{jpconf}
\usepackage{graphicx}
\begin{document}
\title{First ADS analysis of $B^- \to D^0 K^-$ decays in hadron collisions}

\author{Paola Garosi, on behalf of the CDF Collaboration}

\address{INFN Pisa, Largo B. Pontecorvo 3, 56127 Pisa, Italy\\
University of Siena, via Roma 56, 53100 Siena, Italy}

\ead{paola.garosi@pi.infn.it}

\begin{abstract}
The CDF experiment reports the first measurement of branching fractions and CP-violating asymmetries of doubly-Cabibbo suppressed $B^- \to D^0 K^-$ decays in hadron collisions, using the approach proposed by Atwood, Dunietz and Soni (ADS) to determine the CKM angle $\gamma$. Using 5.0 fb$^{-1}$ of data the combined significance of both $B^- \to D^0 \pi/K$ signals exceeds 5 sigma.
First results in hadron collisions, obtained on 1.0 fb$^{-1}$ of data using the method of Gronau, London and Wyler (GLW) for the Cabibbo suppressed modes are also reported.
Both ADS and GLW parameters are determined with accuracy comparable with B factories measurements.

\end{abstract}

\section{Introduction}
\begin{figure}
\begin{center}
\includegraphics[width=2.9in]{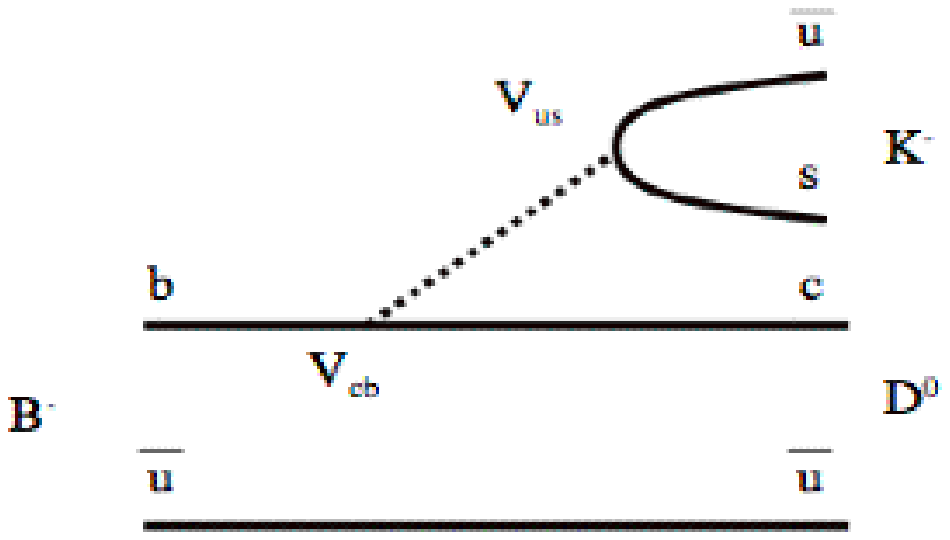}
\includegraphics[width=2.9in]{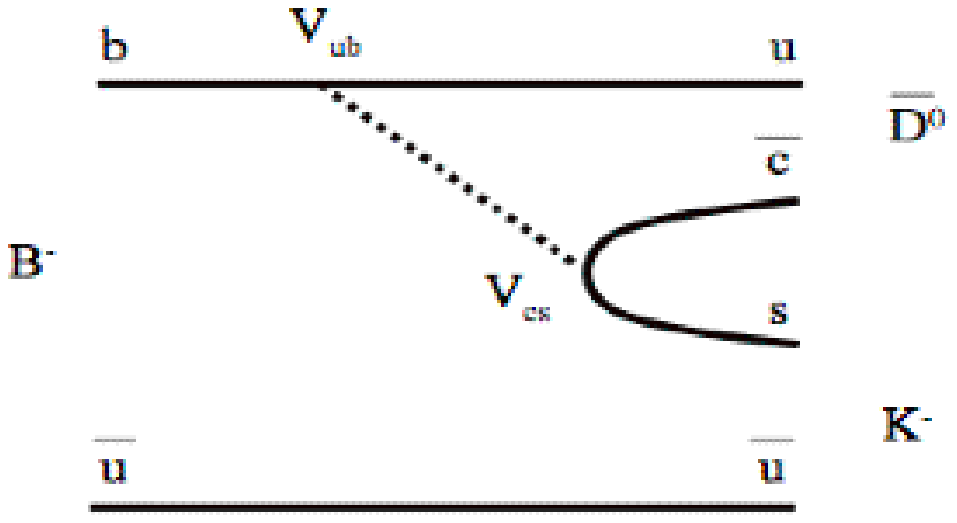}
\end{center}
\caption{\label{fig:diagram}Diagrams contributing to $B \to DK $ modes. On the left the {\it color favored} transition, on the right the {\it color suppressed} transition.}
\end{figure}

The measurement of the CKM matrix elements plays a central role both to test the Standard Model consistency and to probe New Physics scenarios. 
In particular the complex phase of the CKM matrix leads to $CP$ violation in weak processes. Conventionally, $CP$ violating observables are written in terms of the angles $\alpha$, $\beta$ and $\gamma$ of the ``Unitarity Triangle", obtained from the unitarity condition of the CKM matrix~\cite{ref:CKM}.
While the resolution on $\alpha$ and $\beta$ reached a good level of precision, the measurement of $\gamma$ is still limited by the smallness of the branching ratios involved in the processes\cite{ref:hfag,ref:utfit,ref:ckmfitter}.
\\
Written in terms of the CKM elements, $\gamma$ is equal to $arg (- V_{ud} V^*_{ub} / V_{cd} V^*_{cb})$ and is related to the element $V_{ub}$ though $V_{ub} = |V_{ub}| e^{-i \gamma}$~\cite{ref:wolfe}. 
Among the various methods for the $\gamma$ measurement, those which make use of the tree-level $B^- \to D^0 K^-$ decays have the smallest theoretical uncertanties~\cite{ref:glw1,ref:ads1,ref:ggsz}. In fact $\gamma$ appears as the relative weak phase between two amplitudes, the favored $b \to c \bar{u} s$ transition of the $B^- \to D^0 K^-$, whose amplitude is proportional to $V_{cb} V_{us}$, and the color-suppressed $b \to u \bar{c} s$ transition of the $B^- \to \overline{D}^0 K^-$, whose amplitude is proportional to $V_{ub} V_{cs}$. A schematic diagram is shown in Fig.~\ref{fig:diagram}.
The interference between $D^0$ and $\overline{D}^0$, decaying into the same final state, leads to measurable $CP$ violation effects, from which $\gamma$ can be extracted. The effects can be also enhanced choosing the interfering amplitudes of the same order of magnitude.

According to the final state of the $D^0$ we can have the following methods:
\begin{itemize}
\item \textit{GLW (Gronau-London-Wyler) method}~\cite{ref:glw1,ref:glw2}, which uses $CP $ eigenstates of $D^0$, as $D^0_{CP^+} \rightarrow K^+ K^-, \pi^+ \pi^-$ and $D^0_{CP-} \rightarrow K^0_s \pi^0, K^0_s \phi, K^0_s \omega$.
\item \textit{ADS (Atwood-Dunietz-Soni) method}~\cite{ref:ads1,ref:ads2}, which uses the doubly Cabibbo suppressed mode $D^0_{DCS} \rightarrow K^+ \pi^-$.
\item \textit{GGSZ (or Dalitz) method}~\cite{ref:ggsz,ref:ads2}, which uses three body decays of $D^0$, as \\
$D^0 \rightarrow~K^0_s \pi^+ \pi^-$.
\end{itemize} 

All mentioned methods require no tagging or time-dependent 
measurements, and many of them only involve charged particles in 
the final state. They are therefore particularly well-suited to analysis in a hadron 
collider environment, where the large production of $B$ mesons can be well exploited.
The use of specialized trigger based on online detection of secondary vertex (SVT trigger~\cite{ref:trigger}) allows the selection of pure $B$ meson samples.

We will describe in more details the ADS and GLW methods, for which CDF reports the first results in hadron collisions.

\section{CDF II detector and trigger}
The CDF experiment is located at the Tevatron, a $\sqrt{s} = 1.96$ TeV $p\bar{p}$ collider. 
The detector~\cite{ref:cdf1} is a multipurpose magnetic spectrometer surrounded by calorimeters and muon detectors. 
The most relevant for $B$-physics are the tracking, the particle-identification (PID) detectors and the trigger system. \\
The tracking system provides a determination of the decay point of particles with 15 $\mu$m resolution in the transverse plane using six layers of double-sided silicon-microstrip sensors at radii between 2.5 and 22 cm from the beam. 
A 96-layer drift chamber extending radially from 40 to 140 cm from the beam provides the reconstruction of three-dimensional charged-particles trajectories and excellent transverse momentum resolution, $\sigma_{p_T} / p^2_{T} = 0.1\% $ Gev$/c^2$. Specific ionization measurements in the chamber allow $1.5\sigma$ separation between charged kaons and pions, approximately constant at momenta larger than 2 GeV$/c$.
\\
A three-level trigger system~\cite{ref:trigger} selects events enriched in decays of long-lived particles by exploiting the presence of displaced tracks in the event and measuring their impact parameter with offline-like 30 $\mu$m resolution. The trigger requires the presence of two charged particles with transverse momenta greater than 2 GeV$/c$, impact parameters greater than 100 microns and basic cuts on azimuthal separation and scalar sum of momenta.

\section{The Atwood-Dunietz-Soni method}
The ADS method~\cite{ref:ads1,ref:ads2} takes into account the following decay channels:
$B^- \to D^0 K^-$ (\textit{color favored}), with $D^0 \to K^+ \pi^-$ (\textit{doubly Cabibbo suppressed})
and $B^- \to \overline{D}^0 K^-$ (\textit{color suppressed}), with $ \overline{D}^0 \to K^+ \pi^-$ (\textit{Cabibbo favored}).
\\
The final state $[K^+ \pi^-]_D K^-$ is the same and since $D^0$ and $\overline{D}^0$ are undistinguishable, the direct $CP $ asymmetry can be measured.
The interfering amplitudes are of the same order of magnitude, so large asymmetry effects are expected.
For simplicity we will call ``DCS" the final state $[K^+ \pi^-]_D K^-$ and we will use the label $B \to D^0_{DCS} K$ to identify it.

The direct CP asymmetry
$$
\displaystyle A_{ADS}  = \frac{\mathcal{B}(B^-\rightarrow [K^+\pi^-]_{D}K^-)-\mathcal{B}(B^+\rightarrow [K^-\pi^+]_{D}K^+)}{\mathcal{B}(B^-\rightarrow [K^+\pi^-]_{D}K^-)+\mathcal{B}(B^+\rightarrow [K^-\pi^+]_{D}K^+)}
$$ 
can be written in terms of the decay amplitudes and phases:
$$
A_{ADS}  =  \frac{2r_B r_D\sin{\gamma}\sin{(\delta_B+\delta_D)}}{r_D^2 + r_B^2 + 2r_Dr_B \cos{\gamma}\cos{(\delta_B+\delta_D)}}, 
$$ 
where $r_B = |A(b\to u)/A(b\to c)|$, $\delta_B = arg[A(b\to u)/A(b\to c)]$ and $r_D$ and $\delta_D$ are the corresponding amplitude ratio and strong phase difference of the $D$ meson.\\
The denominator corresponds to another physical observable, the ratio between DCS and Cabibbo favored (``CF") events, the latter coming from the decay channel
$B^- \to D^0 K^- $ (\textit{color favored}), with $D^0 \to K^- \pi^+$ (\textit{Cabibbo favored}):
\begin{eqnarray}
R_{ADS}  & = & r_D^2 + r_B^2 + 2r_Dr_B \cos{\gamma}\cos{(\delta_B+\delta_D)} =  \nonumber \\
 & = &   \frac{\mathcal{B}(B^-\rightarrow [K^+ \pi^-]_{D}K^-)+\mathcal{B}(B^+\rightarrow [K^-\pi^+]_{D}K^+)}{\mathcal{B}(B^-\rightarrow [K^- \pi^+]_{D}K^-)+\mathcal{B}(B^+\rightarrow [K^+\pi^-]_{D}K^+)}. \nonumber
\end{eqnarray}
%
%
We can measure the corresponding quantities, $A_{ADS}$ and $R_{ADS}$, also for the 
$B^- \to D^0 \pi^-$ mode, for which sizeable asymmetries may be found~\cite{ref:hfag}.

\begin{figure}[!h]
\centering
\includegraphics[width=3.1in]{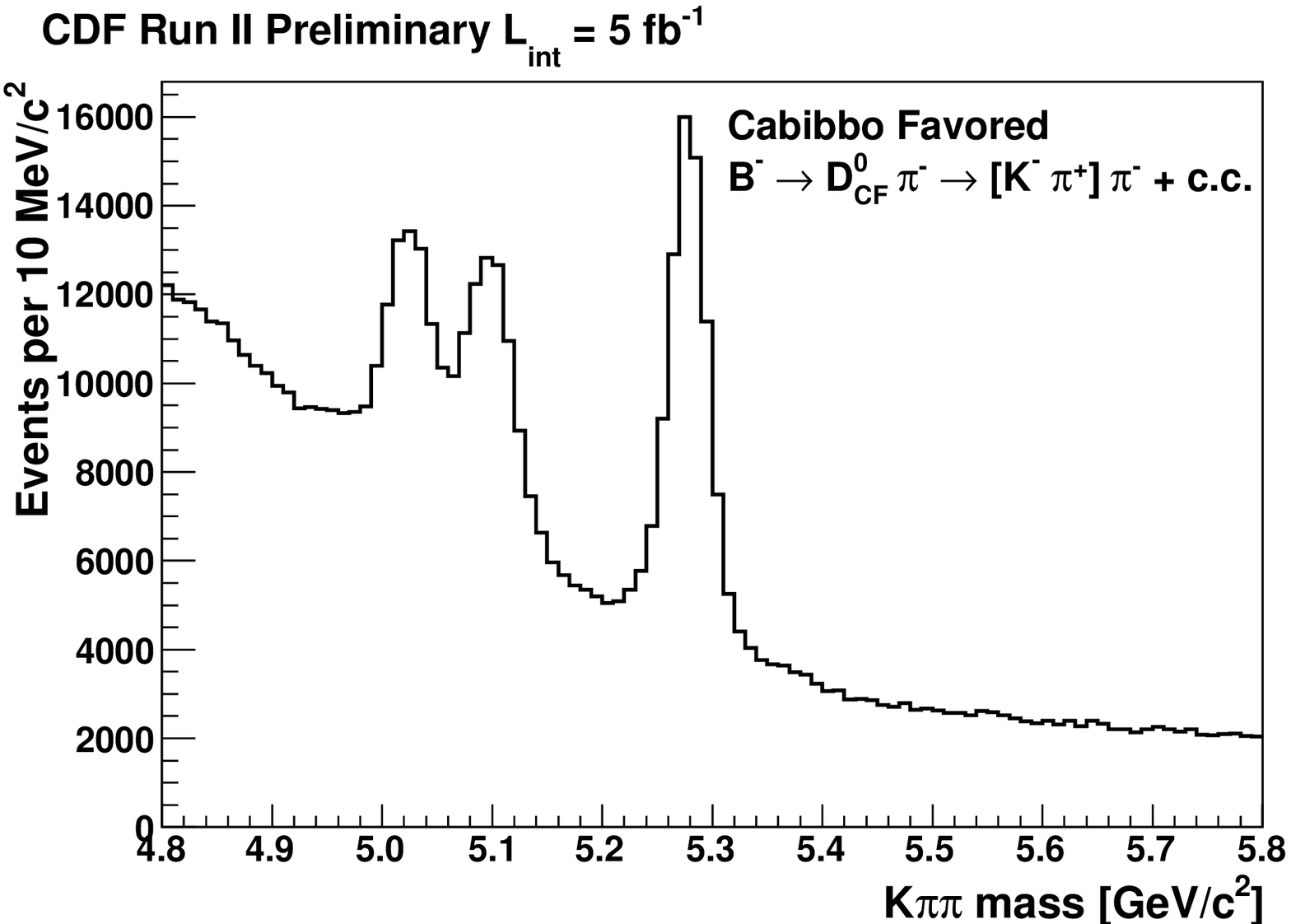} 
\includegraphics[width=3.1in]{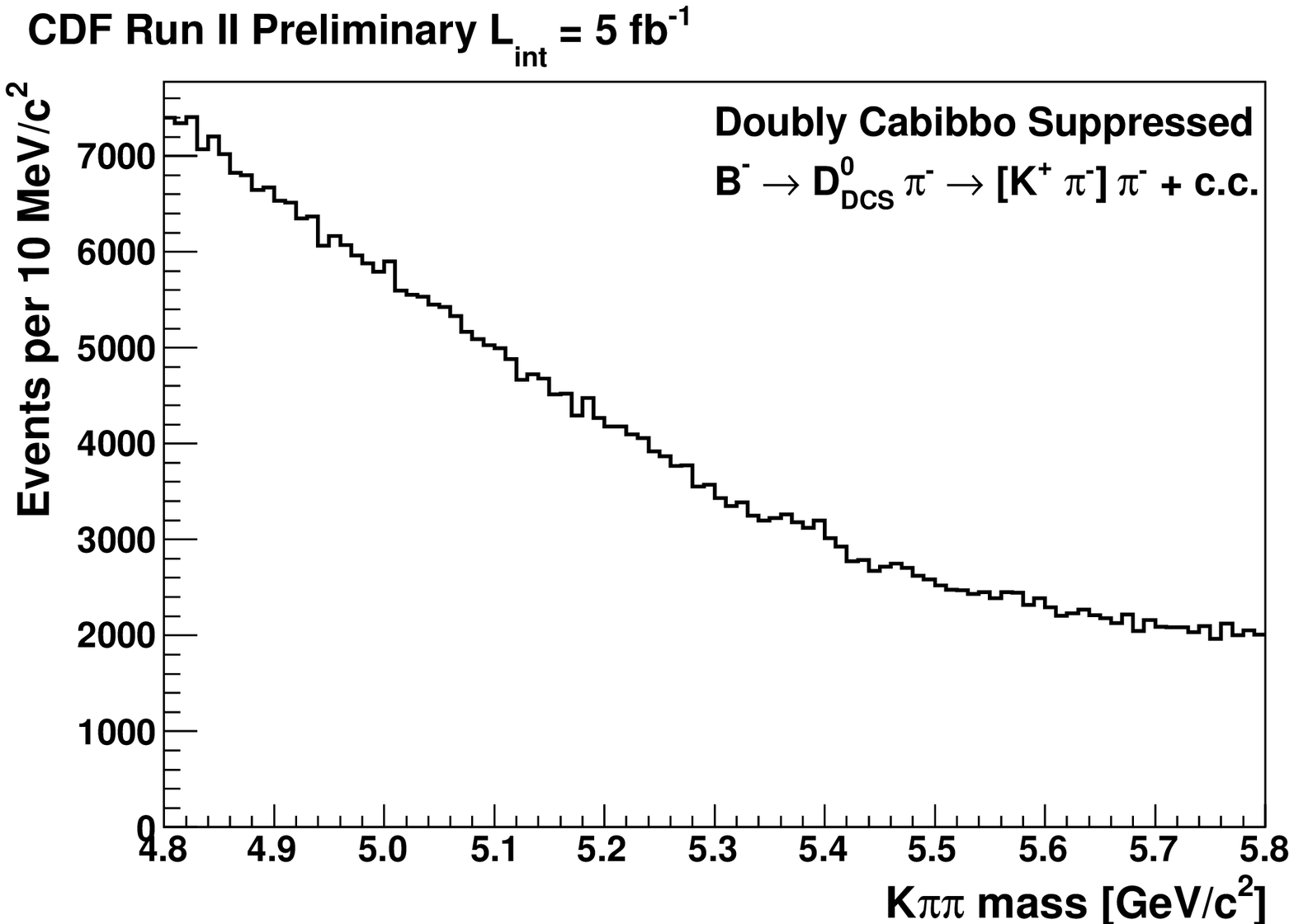}  
\caption{Invariant mass distributions of $B \to D^0 h$ candidates for each reconstructed
decay mode, Cabibbo favored on the left and doubly Cabibbo suppressed on the right. The pion mass is assigned to the track from the B decay.} \label{fig:before_cuts}
\end{figure}

The invariant mass distributions of CF and DCS modes, using a data sample of 5~fb$^{-1}$ of data, with a nominal pion mass assignment to the track from B, are reported in Fig.~\ref{fig:before_cuts}. 

A $B \to D^0 \pi$ CF signal is
visible at the correct mass of about 5.279 GeV$/c^2$. 
Events from $B \to D^0 K$ decays are expected to form much smaller and wider
peak, located about 50 MeV$/c^2$ below the $B \to D^0 \pi$ peak.

The $B \to D^0 \pi$ and $B \to D^0 K$ DCS signals instead appear to be buried in the combinatorial background. 
For this reason an important point of this analysis is the suppression of the combinatorial background, obtained through a cuts optimization focused on finding a signal of the $B \to D_{DCS} \pi$ mode. 
Since the $B \to D_{CF} \pi$ mode has the same topology of the DCS one, but more statistic, we
did the optimization using signal (S) and background (B) directly from CF data, choosing a set of cuts which maximize the figure of merit $S/(1.5+\sqrt{B})$~\cite{ref:punzi}. 

The \textit{offline cut on the tridimensional vertex quality} $\chi^2_{3D}$ and the $B \ isolation$ are powerful handles among the variables used in the optimization. The first exploit the 3D silicon-tracking to resolve multiple vertices along the beam direction and to reject fake tracks.  It allows a background reduction by a factor of two and has small inefficiency on signal (less than 10\%). The $B$ isolation corresponds to the fraction of momentum carried by the $B$ meson, which is usually greater than the momentum carried by lighter mesons.
Another important cut is on the \textit{decay lenght of the $D^0$ with respect to the B}, which allows to reject most of the $B \to hhh$ backgrounds, where $h$ is either $\pi$ or $K$.  
All variables and threshold values applied are described in~\cite{ref:pubnote}. 

\begin{figure}[!ht]
\centering
\includegraphics[width=3.1in]{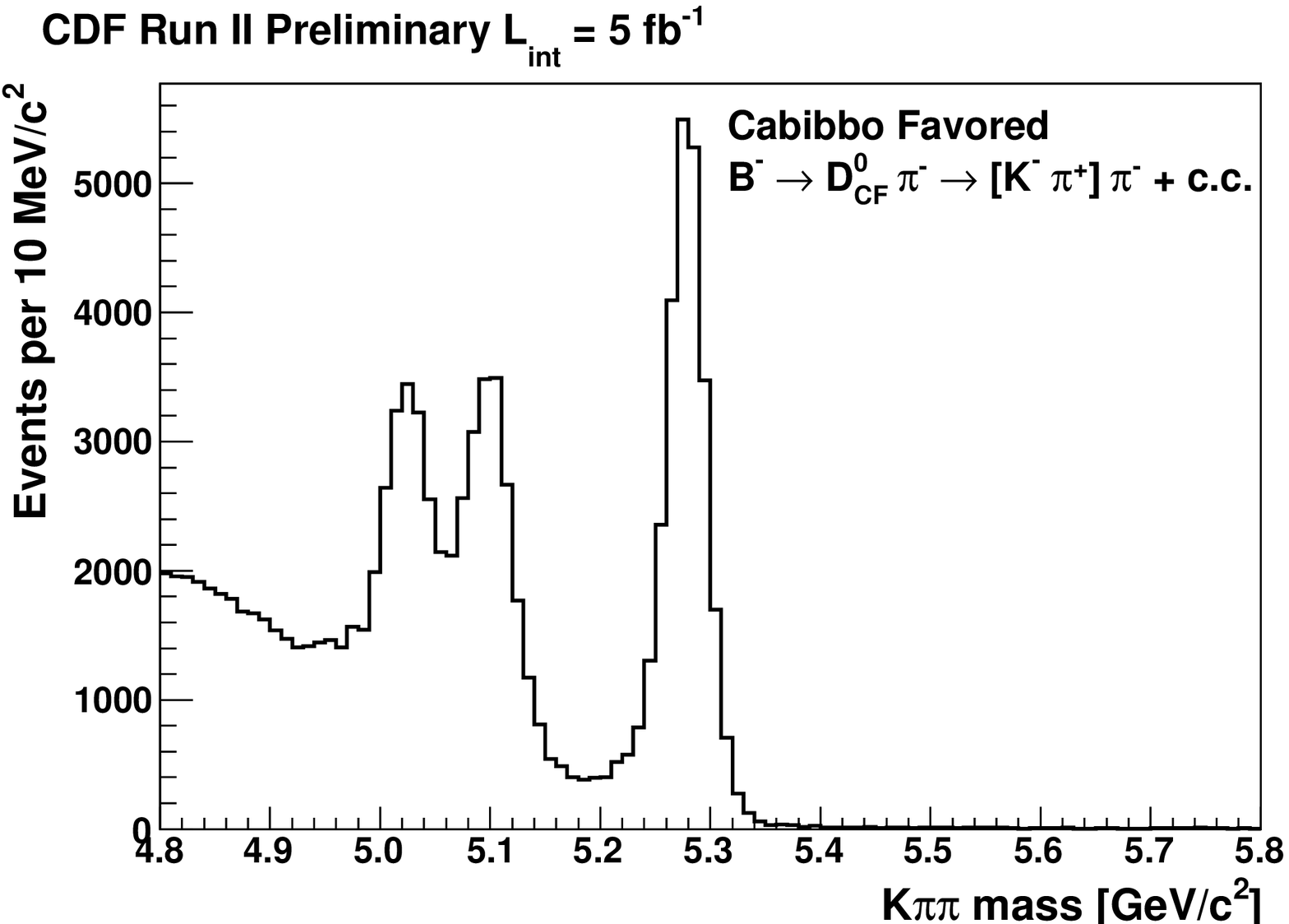}
\includegraphics[width=3.1in]{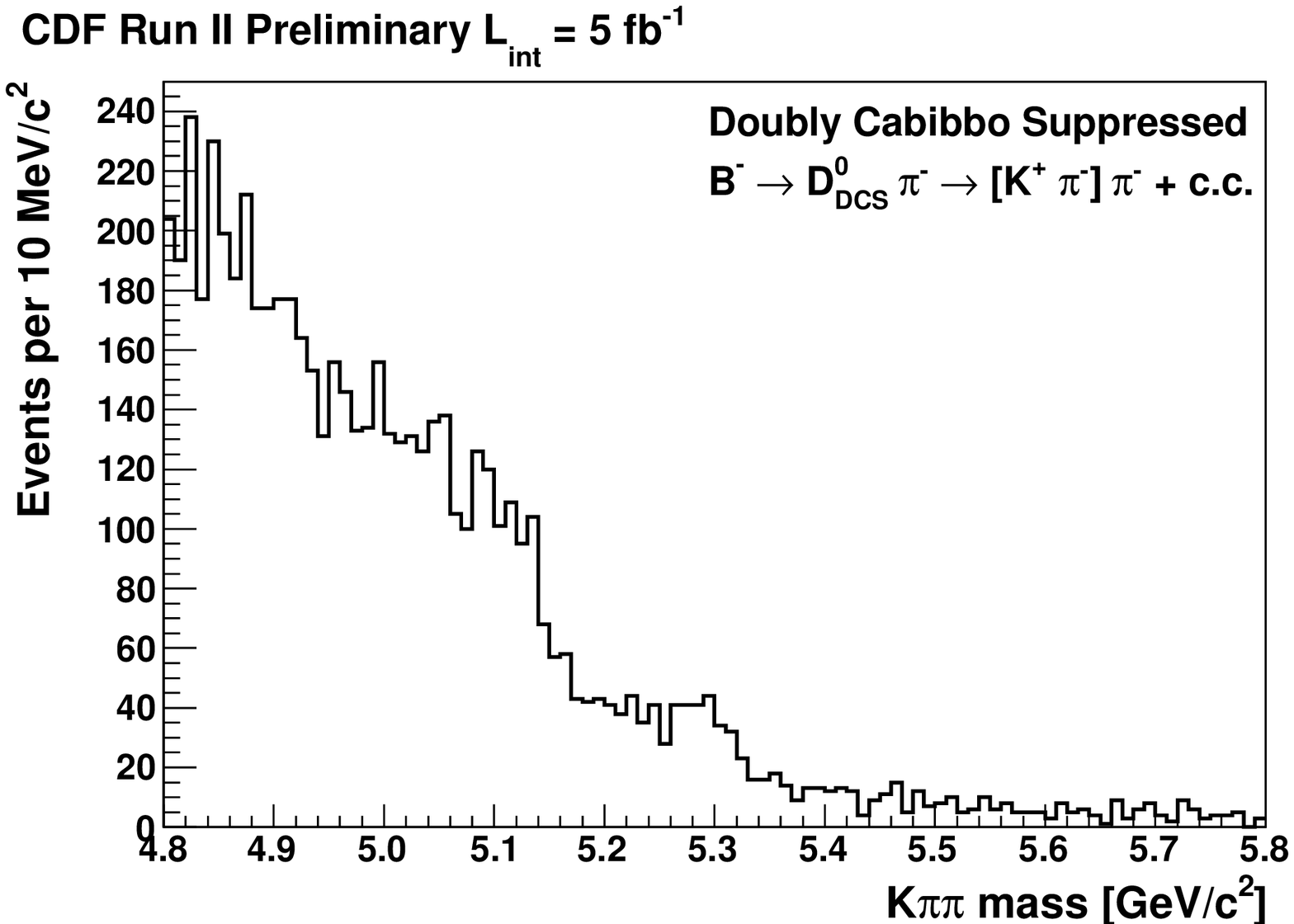} 
\caption{Invariant mass distributions of $B \to D^0 h$ candidates for each reconstructed
decay mode, Cabibbo favored on the left and doubly Cabibbo suppressed on the right, after the cuts optimization. The pion mass is assigned to the track from the B decay.} \label{fig:after_cuts}
\end{figure}

The resulting invariant mass distributions of CF and DCS modes are reported in Fig.~\ref{fig:after_cuts} where the
combinatorial background is almost reduced to zero and an excess of events is now visible in the correct DCS signal mass window.

\begin{figure}[!h]
\centering
\includegraphics[width=3.6in]{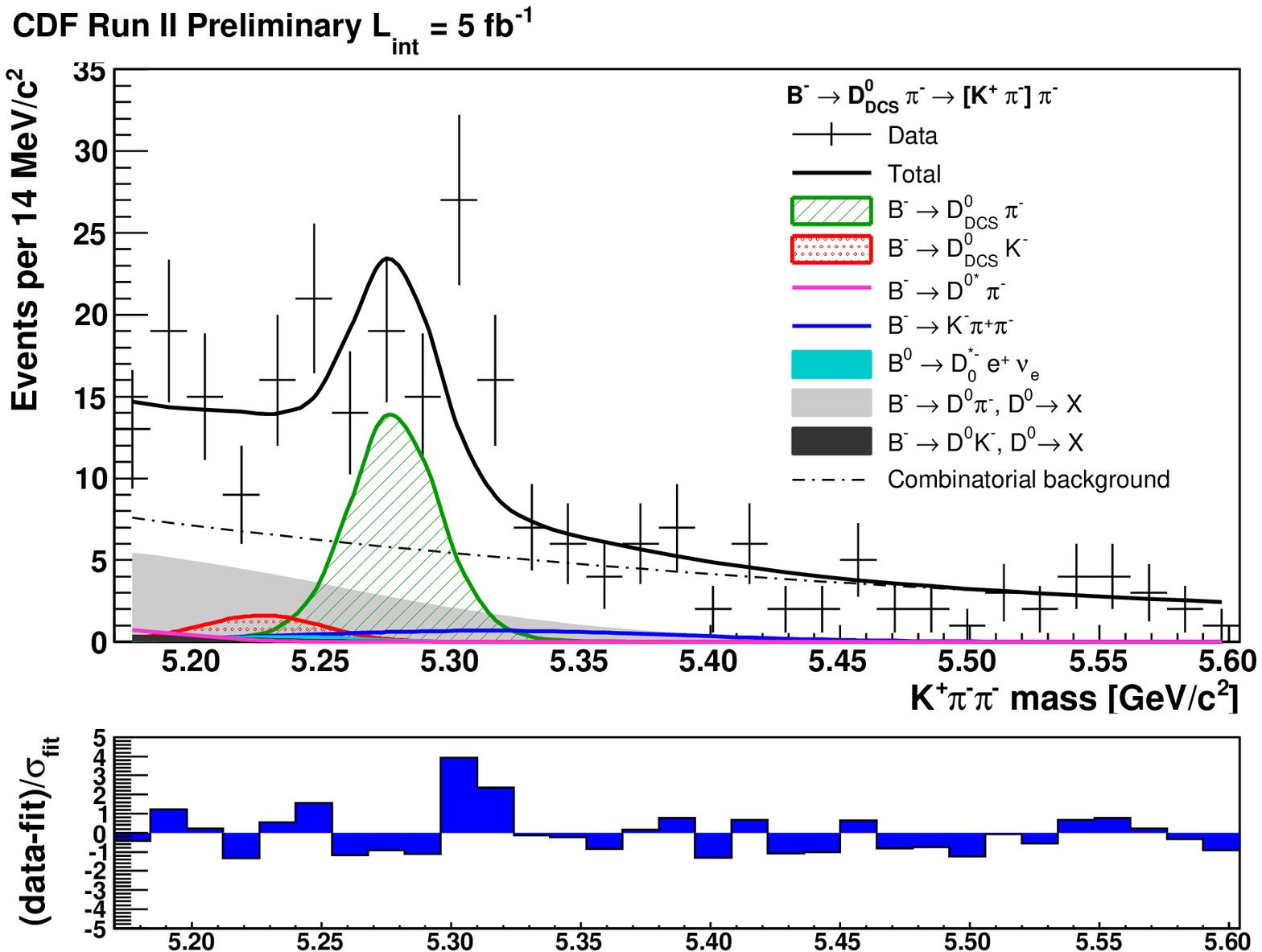} \\  
\vspace{0.5cm}
\includegraphics[width=3.6in]{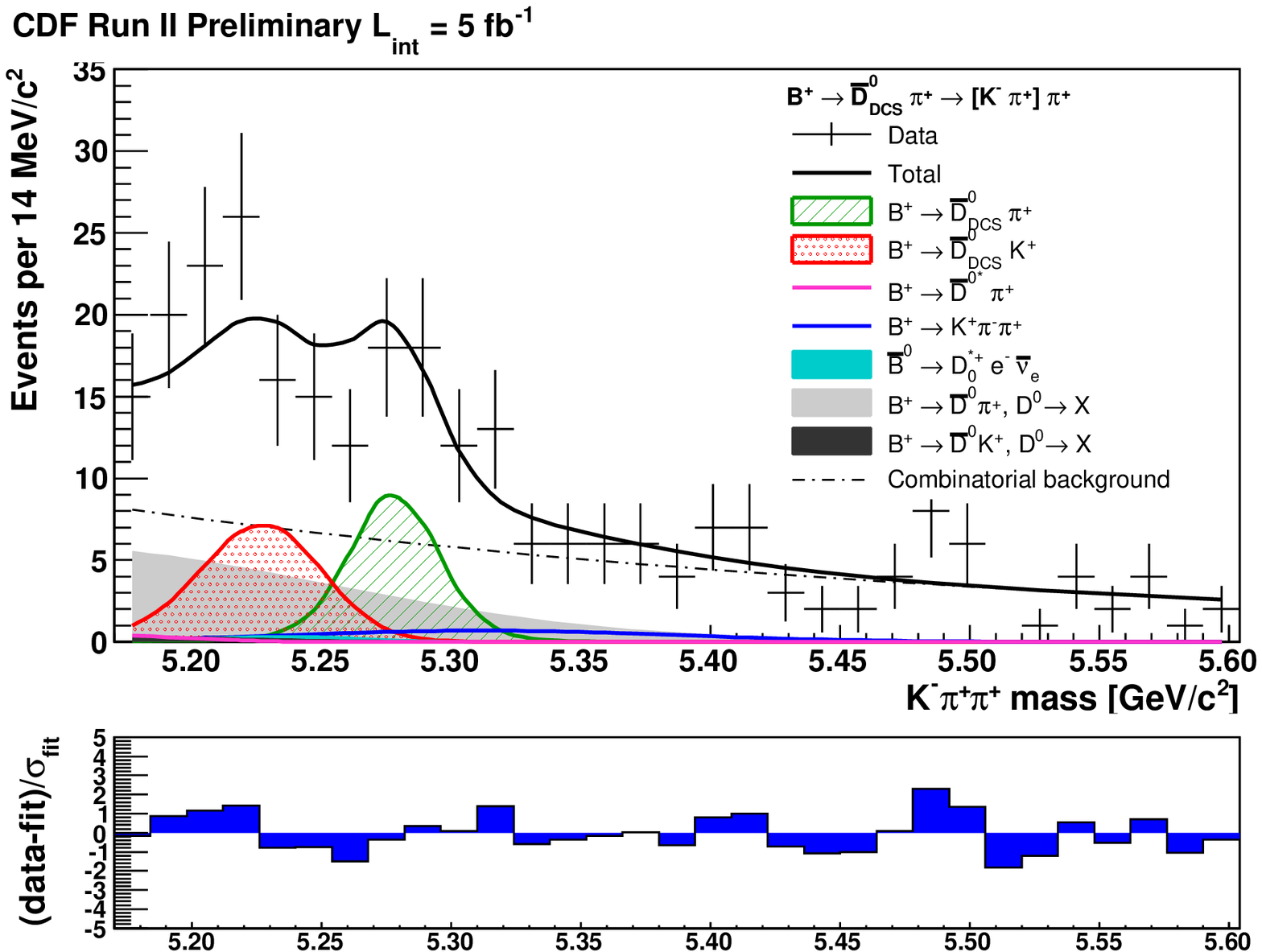} 
\caption{Invariant mass distributions of $B \to D^0_{DCS} h$ candidates for negative (top) and positive (bottom) charges. The pion mass is assigned to the track from the B decay. The projections of the likelihood fit are overlaid.} \label{fig:plot_dcs}
\end{figure}

An unbinned likelihood fit, exploiting mass and particle identification information
provided by the specific ionization (dE$/$dx) in the CDF drift chamber, is performed~\cite{ref:pubnote} 
to separate the $B \to DK $ contributions from the $B \to D \pi$ signals and the combinatorial and physics backgrounds. The dominant physics backgrounds for the DCS mode are $B^- \to D^0 \pi^-$, with $D^0 \to X$; $B^- \to D^0 K^-$, with $D^0 \to X$;	$B^- \to D^{0*} \pi^-$, with $D^{0*} \to D^0 \pi^0/ \gamma$; $B^- \to K^- \pi^+ \pi^-$ and $B^0 \to D^{*-}_0 e^+ \nu_e$.\\
Fig.~\ref{fig:plot_dcs} shows the DCS invariant mass distributions separated in charge. 
\\
We obtained $34 \pm 14$ $ B \to D_{DCS} K $ and $73 \pm 16$ $B \to D_{DCS} \pi$ signal events.\\
Since $K^+$ and $K^-$ have a different probability of interaction in the detector, we evaluated the efficiency using a simulation sample and we corrected the fit results with this value.\\
The final results for the asymmetries are:
\begin{eqnarray}
A_{ADS}(K) & =  & - 0.63 \pm 0.40\mbox{(stat)} \pm 0.23\mbox{(syst)} \nonumber \\
A_{ADS}(\pi) & = &  0.22 \pm 0.18\mbox{(stat)} \pm 0.06\mbox{(syst)} \nonumber
\end{eqnarray}
and for the ratios of doubly Cabibbo suppressed mode to flavor eigenstate: 
\begin{eqnarray}
R_{ADS}(K) & = & [22.5 \pm 8.4\mbox{(stat)} \pm 7.9\mbox{(syst)}] \cdot 10^{-3}  \nonumber \\
R_{ADS}(\pi) & = & [4.1 \pm 0.8\mbox{(stat)} \pm 0.4\mbox{(syst)}] \cdot 10^{-3}.\nonumber
\end{eqnarray}
These quantities are measured for the first time in hadron collisions. The results
are in agreement with existing measurements performed at $\Upsilon$(4S) resonance~\cite{ref:hfag,ref:ckmfitter}.

\section{Gronau-London-Wiler method}
In the GLW method~\cite{ref:glw1,ref:glw2} the CP asymmetry of $B \to D^0_{CP\pm} K$ is studied, where $CP \pm$ are the $CP$ even and odd eigenstates of the $D^0$, as $D^0_{CP^+} \rightarrow K^+ K^-, \pi^+ \pi^-$ and $D^0_{CP-} \rightarrow K^0_s \pi^0, K^0_s \phi, K^0_s \omega$.

We can define four observables: 
\begin{eqnarray}
A_{CP \pm}  & = & \frac{\mathcal{B} (B^- \to D^0_{CP \pm} K^-) - \mathcal{B} (B^+ \to D^0_{CP \pm} K^+)} {\mathcal{B} (B^- \to D^0_{CP \pm} K^-) + \mathcal{B} (B^+ \to D^0_{CP \pm} K^+)} \nonumber \\
R_{CP \pm}  & = & 2 \cdot \frac{\mathcal{B} (B^- \to D^0_{CP \pm} K^-) + \mathcal{B} (B^+ \to D^0_{CP \pm} K^+)} {\mathcal{B} (B^- \to D^0_{CF} K^-) + \mathcal{B} (B^+ \to \overline{D}^0_{CF} K^+)}, \nonumber
\end{eqnarray}
of which only three are indipendent (since $A_{CP+}R_{CP+} = -A_{CP-}R_{CP-}$). 

The relations with the amplitude ratios and phases are: 
\begin{eqnarray}
A_{CP \pm} & = & 2 r_B \sin{\delta_B} \sin{\gamma} / R_{CP \pm} \nonumber \\
R_{CP \pm} & = & 1 + r^2_B \pm   2 r_B \cos{\delta_B} \cos{\gamma}. \nonumber
\end{eqnarray}
The GLW method is very clean, in fact for three independent observables we have three unknowns. Unfortunately the sensitivity to $\gamma$ is  proportional to $r_B$, so we expect to see small asymmetries.

CDF performed the first measurement of branching fraction and $CP$ asymmetry of the $CP+$ modes at a hadron collider, using 1 fb$^{-1}$ of data~\cite{ref:dcp}.

\begin{figure}[!h]
\centering
\includegraphics[width=2.9in]{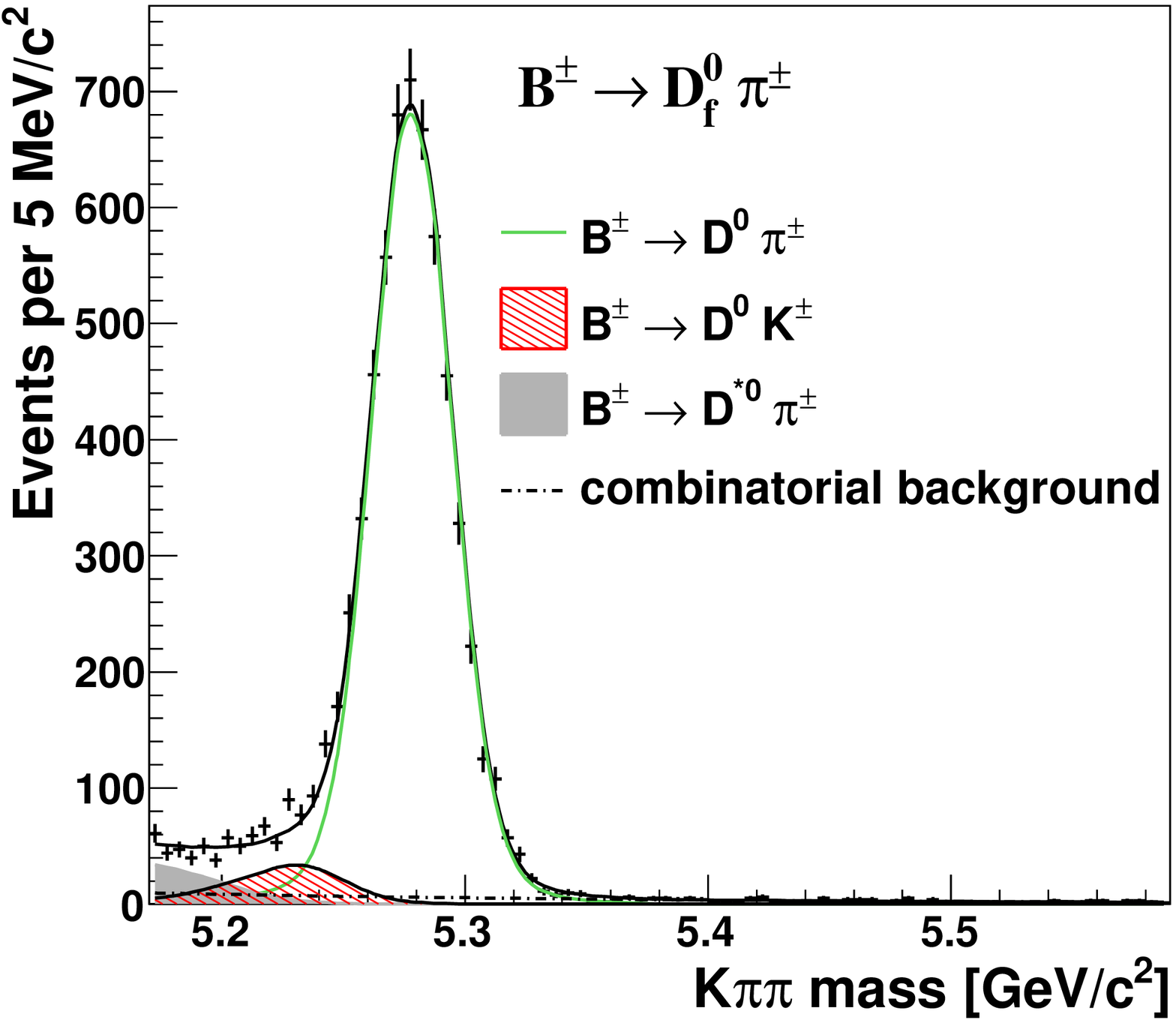}
\includegraphics[width=2.9in]{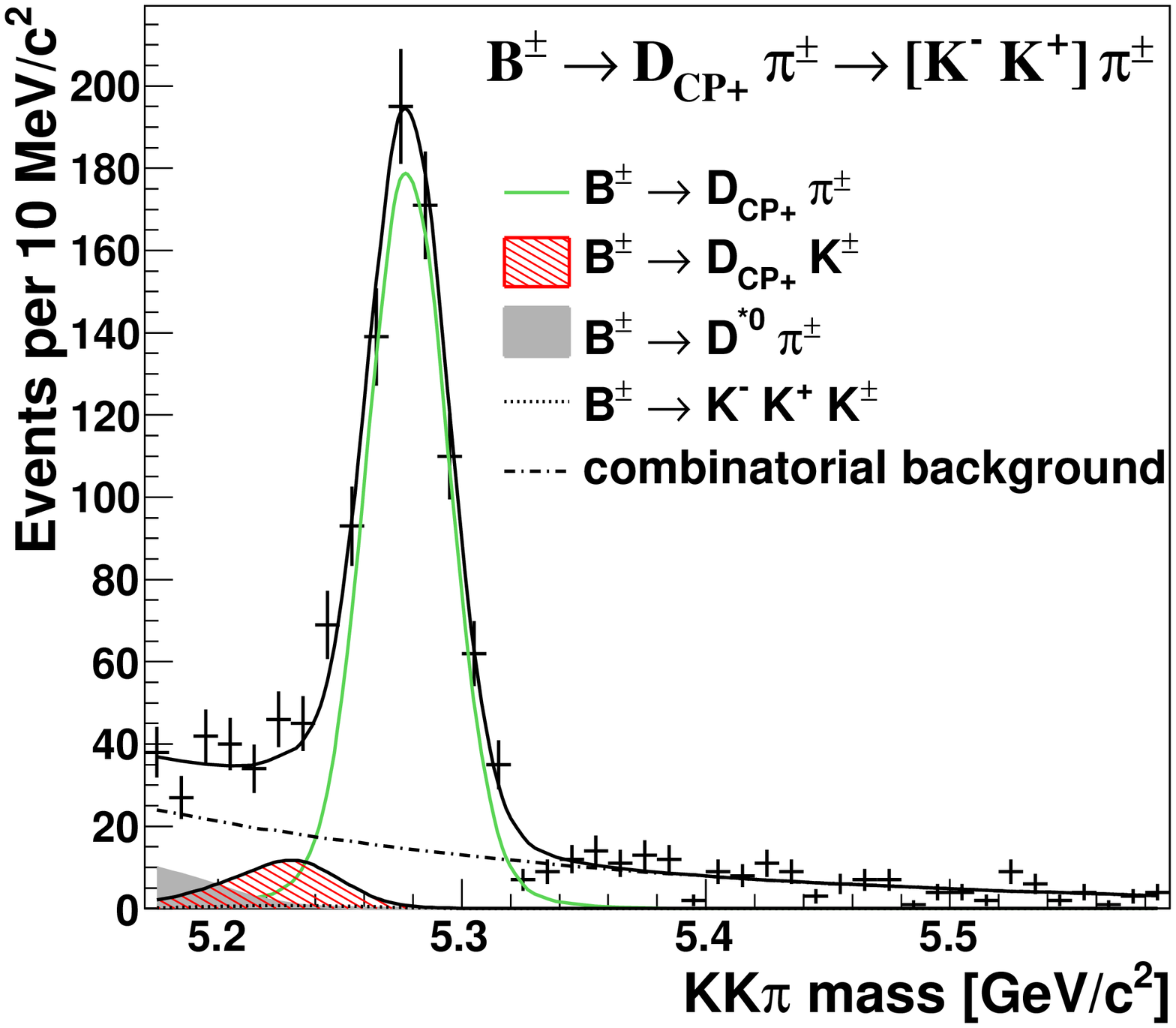}
\includegraphics[width=2.9in]{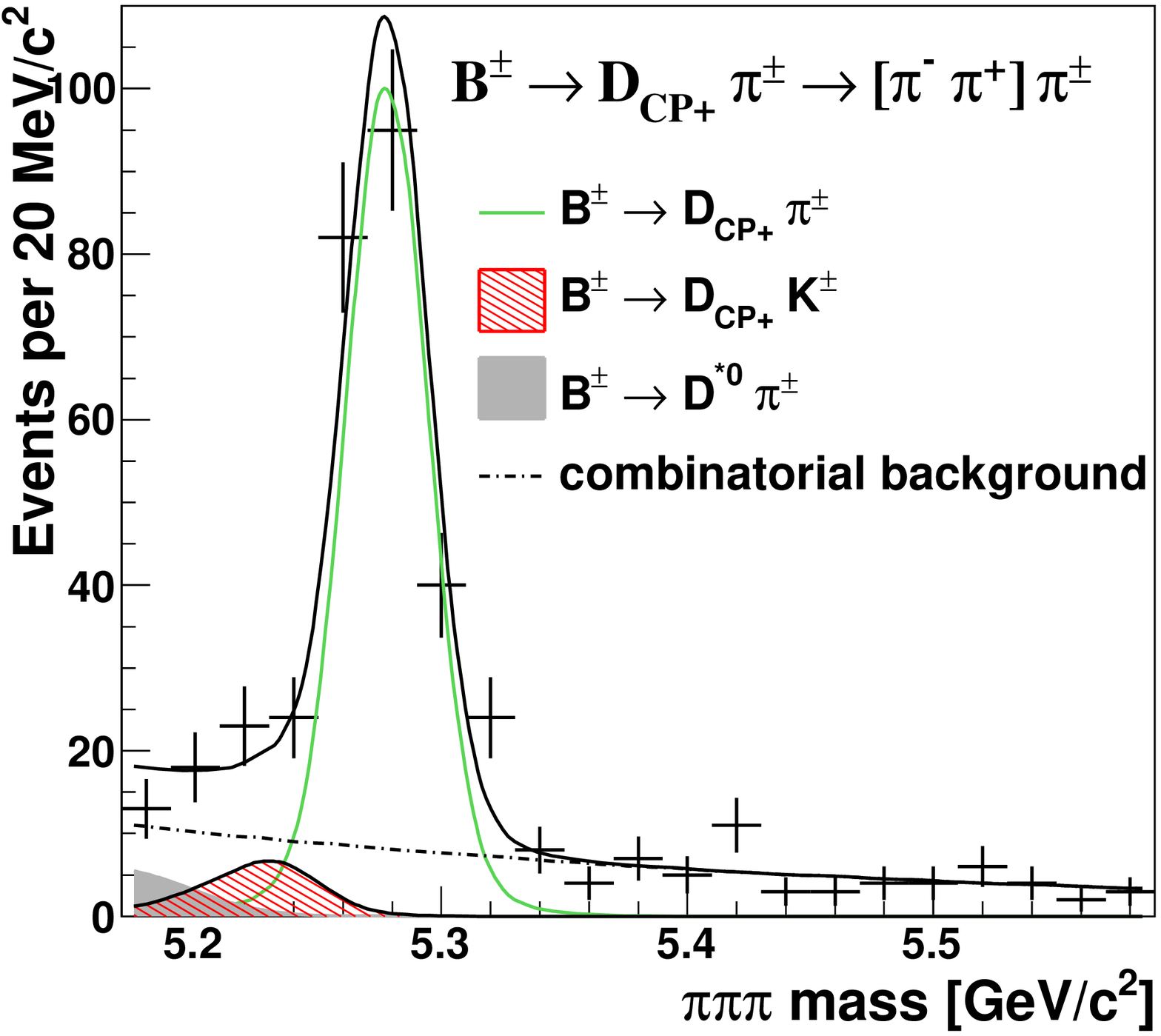} 
\caption{Invariant mass distributions of $B \to D^0 h$ candidates for each reconstructed decay mode, Cabibbo favored on the top left, Cabibbo-suppressed $K^+ K^-$ on the top right and Cabibbo-suppressed $\pi^+ \pi^-$ on the bottom. The pion mass is assigned to the track from the B decay. The projections of the likelihood fit are overlaid for each mode.} \label{fig:plots_glw}
\end{figure}

The mass distributions obtained for the three modes of interest
($D^0 \to K^+\pi^-$, $K^+K^-$ and $\pi^+\pi^-$) are reported in Fig.~\ref{fig:plots_glw}; a clear $B \to D \pi$ signal can be seen
in each plot.

The dominant backgrounds are combinatorial background
and mis-reconstructed physics background such as $B^- \to D^{0*} \pi^-$ decay. In the
$D^0 \to K^+ K^-$ final state also the non resonant $B^- \to K^-K^+K^-$ decay appears, as determined by a study
on CDF simulation~\cite{ref:notaDcp}. 

An unbinned maximum likelihood fit, exploiting kinematic and particle
identification information provided by the dE$/$dx, is performed to statistically separate the $B \to D^0 K$ contributions from the $B \to D^0 \pi$ signals and from the combinatorial and physics backgrounds.

We obtained about 90 $B \to D^0_{CP +} K$ events and we measured the double ratio of CP-even to flavor eigenstate branching fractions 
$$
R_{CP+} = 1.30 \pm 0.24 \mbox{(stat)} \pm  0.12 \mbox{(syst)}
$$
and the direct CP asymmetry 
$$A_{CP+} =  0.39 \pm 0.17 \mbox{(stat)} \pm  0.04 \mbox{(syst)}
$$
These results are in agreement with previous measurements from $\Upsilon$(4S) decays~\cite{ref:hfag,ref:ckmfitter}.

\section{Conclusions}
The CDF experiment is pursuing a global program to measure the $\gamma$ angle from tree-dominated processes. 
The published measurement using the GLW method and the preliminary result using the ADS method show competitive results with previous measurements performed at $B$-factories and demonstrate the feasibility of these kind of measurements also at a hadron collider.
 
We expect to double the data-set available by the end of the year 2010 and obtain interesting and more competitive results in the near future.

\end{document}